\documentclass[twocolumn,10pt]{revtex4} 
\usepackage{amsmath,amssymb,graphicx}

\newcommand{\de}{\partial}

\newcommand{\eq}[2]{\begin{equation} \label{#1} #2 \end{equation}}

\newcommand{\etal}{{\em et al.}}

\newcommand{\schr}{Schr\"odinger }

\begin{document}

\title{Contribution of third-harmonic and negative frequency polarization fields to self-phase modulation in nonlinear media}

\author{Cristian Redondo Loures}\email{Corresponding author: cr169@hw.ac.uk}
\affiliation{School of Engineering and Physics, Heriot-Watt University, EH14 4AS, Edinburgh (UK)}
\author{Andrea Armaroli}
\affiliation{Max Planck Institute for the Science of Light, 91058 Erlangen, Germany}
\author{Fabio Biancalana}
\affiliation{School of Engineering and Physics, Heriot-Watt University, EH14 4AS, Edinburgh (UK)}

\begin{abstract}
We study the influence of third-harmonic generation (THG) and negative frequency polarization terms in the self-phase modulation (SPM) of short and intense pulses in Kerr media. We find that THG induces additional symmetric lobes in the SPM process. The amplitude of these new sidebands are greatly enhanced by the contributions of the negative frequency Kerr (NFK) term and the shock operator. We compare our theoretical predictions based on the analytical nonlinear phase with simulations carried out by using the full unidirectional pulse propagation equation (UPPE).
\end{abstract}

\maketitle 



Self-phase modulation (SPM) is one of the most basic and fundamental effects in nonlinear optics \cite{shimizu,alfano,shenloy,stolenlin,agrawalbook}. When an ultrashort and intense pulse propagates through a nonlinear Kerr medium, its spectrum broadens continuously with the propagation distance, developing characteristic lobes in the frequency space, while the broadening is only limited by losses \cite{agrawalbook}.

SPM is usually considered in the framework of the nonlinear \schr equation (NLSE), which is written under the assumption of a slowly-varying envelope approximation (SVEA), i.e. when using envelopes instead of real oscillating electric fields, and assuming that the bandwidth of the pulse is much smaller than its carrier frequency. The NLSE gives very reasonable results, and is more than adequate under a broad variety of parameters and conditions.
Regarding SPM, the NLSE predicts the formation of lobes in the spectrum, the detuning of which from the central frequency increases linearly with the propagation distance. This behaviour is very well understood and the theory for this conventional SPM is well known in the literature \cite{stolenlin,alfano}. The nonlinear phase that accumulates during propagation has a very simple form, which depends on the specific shape of the input pulse, and it is entirely due to the Kerr term in the NLSE.

In this Letter we show that for sufficiently short and intense pulses, additional effects must be included in the equations, namely the third-harmonic generation (THG) and the negative frequency Kerr term (first described in Ref. \cite{conforti}). These terms in the nonlinear polarisation are naturally included in the nonlinear Maxwell equations, but they are usually discarded in the NLSE. However they play an important role in the formation of additional sidebands in the SPM process, which we describe here for the first time.


In a recent work, which was based on previous experimental and theoretical efforts \cite{rubino1,conforti2}, Conforti \etal\ \cite{conforti} have derived a new propagation equation that is able to accurately describe the dynamics of ultrashort pulses in non-linear media:
\eq{eq1}{i\de_{\xi}A+\frac{1}{2}s\de_{\tau}^{2}A+\hat{S}(i\de_{\tau})\left[|A|^{2}A+|A|^{2}A^{*}e^{2i\phi}+\frac{1}{3}A^{3}e^{-2i\phi}\right]_{+}=0,}
where $A$ in the envelope of the {\em analytic signal} of the full electric field $E(\xi,\tau)$ [i.e. the envelope of the positive-frequency part of $E$], $\xi$ and $\tau$ are the dimensionless spatial and temporal variables, $\phi(\xi,\tau)\equiv \kappa\xi+\mu\tau$, $\kappa\equiv(\beta_{1}\omega_{0}-\beta_{0})L_{\rm NL}$ is a quantity that is proportional to the difference between the inverse group velocity $\beta_{1}$ and the inverse phase velocity $\beta_{0}/\omega_{0}$ in the medium, $\mu\equiv\omega_{0}t_{0}$, $\omega_{0}$ is the central pump frequency, $t_{0}$ is the input pulse duration, $L_{\rm NL}\equiv(\omega_{0}n_{2}I_{0}/c)^{-1}$ is the nonlinear length, $n_{2}$ is the nonlinear refractive index of the medium, $I_{0}$ is the input intensity of the pulse, and $c$ is the speed of light in vacuum. The scaling of the physical space and time variables is given by $z=\xi L_{\rm NL}$ and $t=\tau t_{0}$, while the dimensionless field $A$ is scaled in such a way that $|A|^{2}I_{0}$ is the physical intensity of the pulse. The coefficient $s$ in Eq. (\ref{eq1}) is $\pm 1$ depending whether one has anomalous or normal group velocity dispersion respectively. The operator $\hat{S}(i\de_{\tau})\equiv\left(1+\frac{i}{\mu}\de_{\tau}\right)$ that appears in Eq. (\ref{eq1}) before the nonlinear terms is the shock operator, which is responsible for the self-steepening of pulses. 

Equation (\ref{eq1}) is written for the envelope of the analytic signal of the electric field (see Ref. \cite{amir} for details), and does not suffer from the limitations of conventional envelope-based approaches like the NLSE, i.e. does not require a SVEA, see Ref. \cite{conforti} for more explanations. Note that an important ingredient in Eq. (\ref{eq1}) is the spectral filtering operation (indicated with a subscript $+$ in the non-linear part), which truncates all the negative frequencies that appear due to the non-linear interactions \cite{conforti}. This truncation is essential in dealing with sub-cycle pulses, for which the SVEA is completely invalid. However, for pulses of a few hundreds of femtoseconds this filtering operation is not strictly essential and it leads only to small differences with the unfiltered equation. For increasing pulse durations, the envelope of the analytic signal resembles more and more the conventional SVEA envelope.

The first nonlinear term in square brackets  in Eq. (\ref{eq1}) is the usual Kerr term, which is responsible of the conventional SPM lobes described many times in various works \cite{shimizu,alfano,shenloy,stolenlin,agrawalbook}. The third nonlinear term in Eq. (\ref{eq1}) is the THG term. The second nonlinear term is a novel one [we dub it {\em negative frequency Kerr} (NFK)  term], which was discovered in Ref. \cite{conforti}, and is mainly responsible for the contribution of negative frequencies in the nonlinear polarization.

Equation (\ref{eq1}) is completely equivalent to the well-known unidirectional pulse propagation equation (UPPE) \cite{UPPE}, which is a forward Maxwell equation which uses a full electric field formulation (see also Refs. \cite{kinsler,taflove} for general reviews on the subject). The main difference between the UPPE and our Eq. (\ref{eq1}) is that the latter is analytically tractable, easier to simulate and physically more transparent, but the two equations have identical physical content.


In order to study the SPM process in Eq. (\ref{eq1}), we neglect the linear dispersion (i.e. we put $s=0$), since we are interested in the strongly nonlinear regime, where the nonlinear length $L_{\rm NL}$ is much shorter than the dispersive length $L_{\rm D2}\equiv t_{0}^{2}/|\beta_{2}|$, where $\beta_{2}$ is the second order dispersion coefficient calculated at $\omega_{0}$. We also neglect the effect of the shock term, although strictly speaking the latter is important for a precise energy conservation, see Ref. \cite{conforti}. As mentioned above, we also do not apply the filtering operation, since the pulses we consider here are well above the subcycle condition. With the above approximations, Eq. (\ref{eq1}) becomes the following ordinary differential equation:
\eq{eq2}{\frac{dA}{d\xi}=i\left[|A|^{2}A+|A|^{2}A^{*}e^{2i\phi}+\frac{1}{3}A^{3}e^{-2i\phi}\right].}


Following a standard procedure explained in Ref. \cite{agrawalbook}, we write the complex envelope in the form $A(\xi,\tau)=V(\xi,\tau)e^{i\phi_{\rm NL}(\xi,\tau)}$, where $V$ and $\phi_{\rm NL}$ are two real fields that correspond to the amplitude and the non-linear phase respectively. Inserting this definition into Eq. (\ref{eq2}), we obtain the following two coupled ordinary differential equations for $V$ and $\phi_{\rm NL}$:
\begin{eqnarray}
\frac{dV}{d\xi}+\frac{1}{3}\left(3\alpha_{\rm NFK}-\alpha_{\rm THG}\right)V^{3}\sin\left[2(\phi-\phi_{\rm NL})\right]=0, \label{ode1} \\
\frac{d\phi_{\rm NL}}{d\xi}=V^{2}\left\{ 1+\left(\alpha_{\rm NFK}+\frac{1}{3}\alpha_{\rm THG}\right)\cos\left[2(\phi-\phi_{\rm NL})\right] \right\}.\label{ode2}
\end{eqnarray}
In Eqs. (\ref{ode1}-\ref{ode2}), $\alpha_{\rm NFK}$ and $\alpha_{\rm THG}$ are two coefficients that can take the values $0$ or $1$, depending which nonlinear terms one wishes to activate or not. Furthermore, we define $\alpha\equiv\alpha_{\rm NFK}+\alpha_{\rm THG}/3$, which is the coefficient that multiplies the cosine in Eq. (\ref{ode2}). This coefficient turns out to be crucial in the following discussions. In the case when one has THG only, with NFK effect removed, one has $\alpha=1/3$. When the NFK is included, together with the THG, one has $\alpha=4/3$. When neither effects are included, but one only retains the Kerr effect, we have $\alpha=0$.


Equations (\ref{ode1}-\ref{ode2}) are strongly nonlinear and a full analytical solution is not available. However, it is possible to find a fully analytical solution in the case where V is independent on $\xi$. This is clearly an approximation, since due to the presence of the THG and NFK terms, the amplitude $V$ must change with $\xi$, which is easily seen by looking at Eq. (\ref{ode1}). Only in the case $\alpha_{\rm NFK}=\alpha_{\rm THG}=0$, one can have exactly $dV(\xi,\tau)/d\xi=0$. However, we have checked numerically that $dV(\xi,\tau)/d\xi\simeq0$ is a good approximation even in the case of $\alpha_{\rm NFK,THG}\neq 0$, since this results only in small oscillations on the body of the pulse, due to third- and higher-harmonics generation during propagation. These oscillations are rather small when compared to the pulse profile, and the pulse maintains its overall shape if the propagation distances are not too long. Therefore we assume for the moment that $V(\tau)$ is the initial pulse profile and does not change during the spatial propagation.

Under the above assumption, the full analytical solution of Eq. (\ref{ode2}) is given by
\eq{phinl1}{\phi_{\rm NL}(\xi,\tau)=\kappa\xi+\mu\tau-\arctan\left\{\frac{\rho_{1}\tan\left(\Gamma\xi+\arctan\left[\frac{\rho_{2}\tan(\mu\tau)}{\Gamma}\right]\right)}{\Gamma}\right\},}
where $\rho_{1}(\xi,\tau)\equiv \kappa-V^{2}(1+\alpha)$, $\rho_{2}(\xi,\tau)\equiv V^{2}(\alpha-1)+\kappa$, $\Gamma(\xi,\tau)\equiv\sqrt{\rho_{1}\rho_{2}}=\sqrt{V^{4}(1-\alpha^{2})+\kappa(\kappa-2V^{2})}$. For typical physically relevant values, $\rho_{1,2}$ and $\Gamma$ are positive and real quantities.

The nonlinear phase of Eq. (\ref{phinl1}) could be used directly in the definition of the envelope $A=Ve^{i\phi_{\rm NL}}$ to obtain its spectrum for different values of propagation distances $\xi$, in order to track the evolution of the SPM lobes. We use Eq. (\ref{phinl1}) for our calculations, however a useful approximation for this complicated phase can be found by expanding Eq. (\ref{phinl1})  in Taylor series around $\alpha=0$:
\eq{phinl2}{\phi'_{\rm NL}\simeq V^{2}\xi-\alpha\frac{V^{2}}{2(\kappa-V^{2})}\left\{\sin(2\mu\tau)-\sin\left[2\mu\tau+2(\kappa-V^{2})\xi\right]\right\}.}
In the form of Eq. (\ref{phinl2}), it is immediately visible that, for $\alpha\rightarrow0$, one recovers the 'old' and very well-known non-linear phase $\phi_{\rm NL}=V^{2}\xi$ which is responsible of the conventional SPM sidebands and lobes. The two functions $\phi'_{\rm NL}$ and $\phi_{\rm NL}$ are almost identical even for $\alpha=4/3$. The second term in Eq. (\ref{phinl2}), proportional to $\alpha$, is the part of the nonlinear phase that is responsible for the new sidebands that we describe for the first time in this Letter.

\begin{figure}[h!]
\centerline{\includegraphics[width=1\columnwidth]{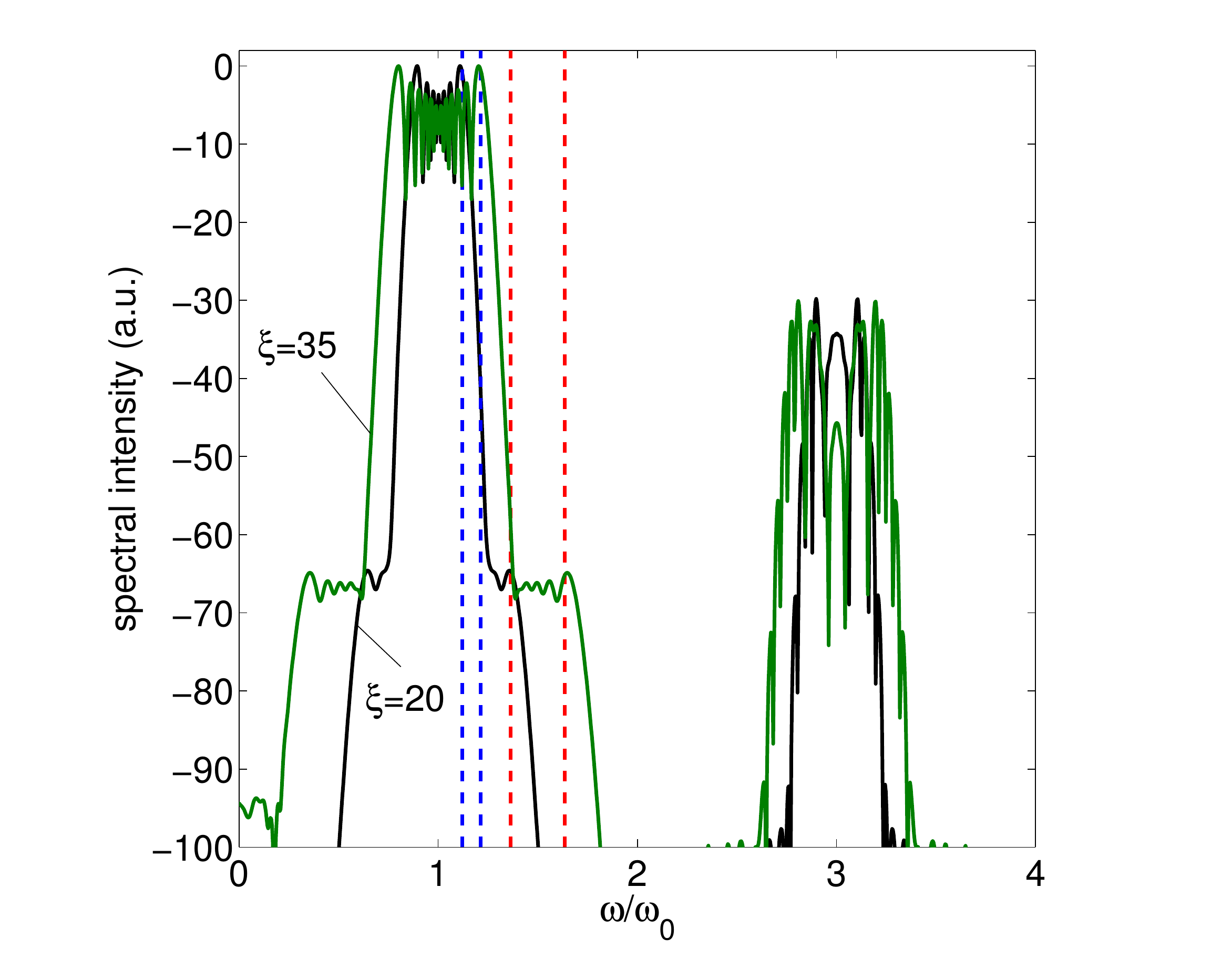}}
\caption{Spectral evolution of a Gaussian pulse for $\xi=20$ and $\xi=35$ in logarithmic scale. Parameters are $\alpha=4/3$, $V_{0}=1$, $\kappa=13.8$ and $\mu=200$. The dashed vertical lines are the predictions for the positions of the conventional SPM lobes and the new lobes.} \label{fig1}
\end{figure}

We now use the analytical expression for the nonlinear phase Eq. (\ref{phinl1}), and we plot the spectral evolution of a Gaussian pulse $V=V_{0}e^{-\tau^{2}/2}$ for different values of $\xi$, in logarithmic scale. Fig. \ref{fig1} shows the evolution of the SPM lobes for $\xi=20$ and $\xi=35$ for physically relevant values of $V_{0}$, $\kappa$ and $\mu$, with $\alpha=4/3$, which is the value that one obtains when including both THG and NFK terms. One notices immediately the presence of new sidebands in the spectra, smaller than the main SPM lobes, but visible and relatively far detuned from the pump frequency. These sidebands are the new feature that we introduce in this Letter.

\begin{figure}[h!]
\centerline{\includegraphics[width=1\columnwidth]{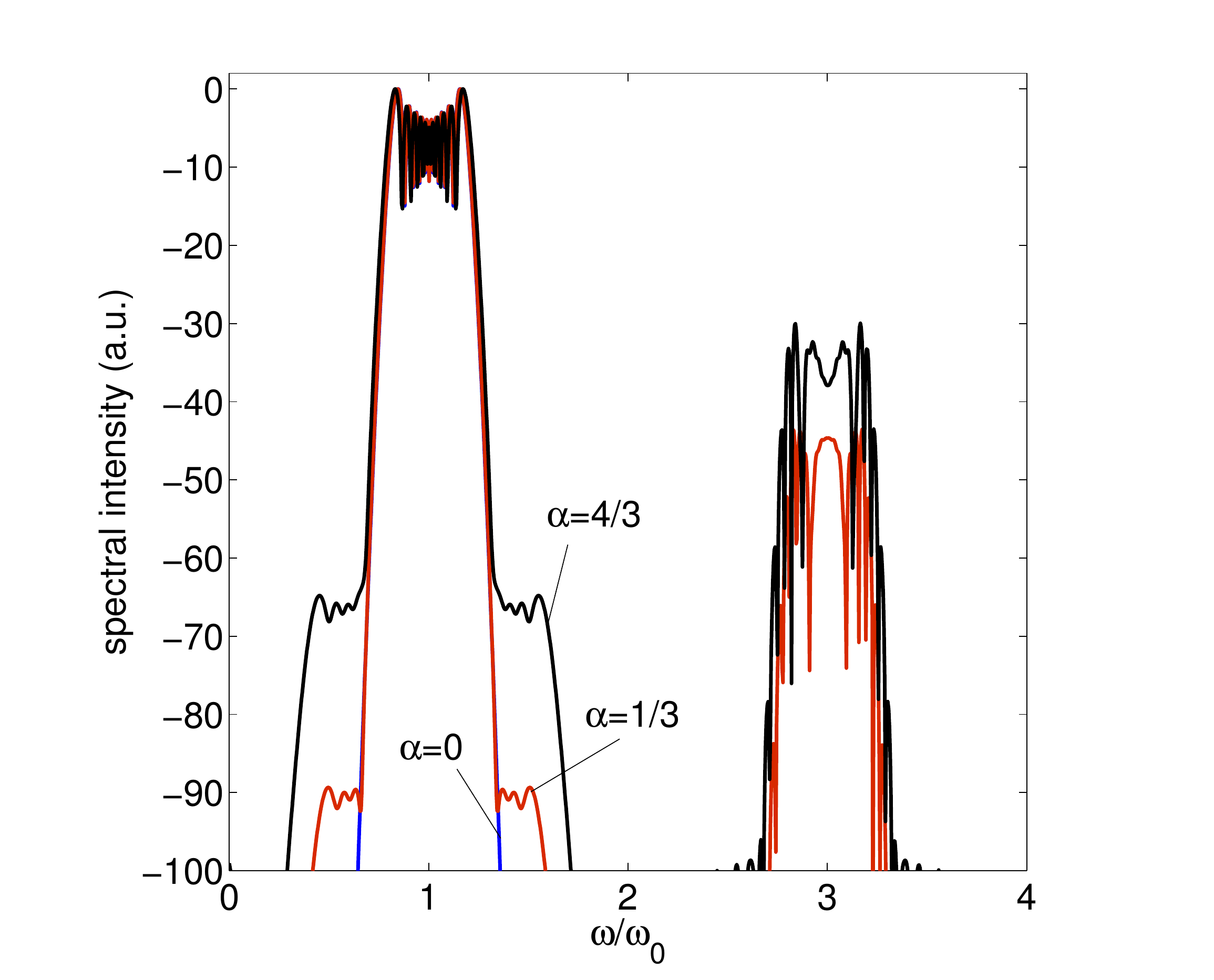}}
\caption{Comparison between different final spectra for the cases $\alpha=0$ (no THG nor NFK), $\alpha=1/3$ (THG only but no NFK) and $\alpha=4/3$ (both THG and NFK included), for $\xi=30$. It is easy to see that the NFK term greatly enhances the amplitude of the additional sidebands. Other parameters are the same as in Fig. \ref{fig1}.} \label{fig2}
\end{figure}

In Fig. \ref{fig2} we show the spectra obtained for $\alpha=0$ (only Kerr term included), $\alpha=1/3$ (Kerr + THG, but no NFK), and $\alpha=4/3$ (all terms included), for a fixed propagation length $\xi=30$. One can immediately notice that the actual position of the additional lobes of SPM is similar in the two cases of $\alpha=1/3$ and $\alpha=4/3$, and so it is approximately independent on the value of $\alpha$. However, the amplitudes are dramatically different, and the NFK term enhances the amplitude of several orders of magnitude. Therefore, we can conclude that, even though the THG is the essential ingredient in the formation of the new sidebands, the NFK greatly contributes to increase the emission of this radiation.

\begin{figure}[h!]
\centerline{\includegraphics[width=1\columnwidth]{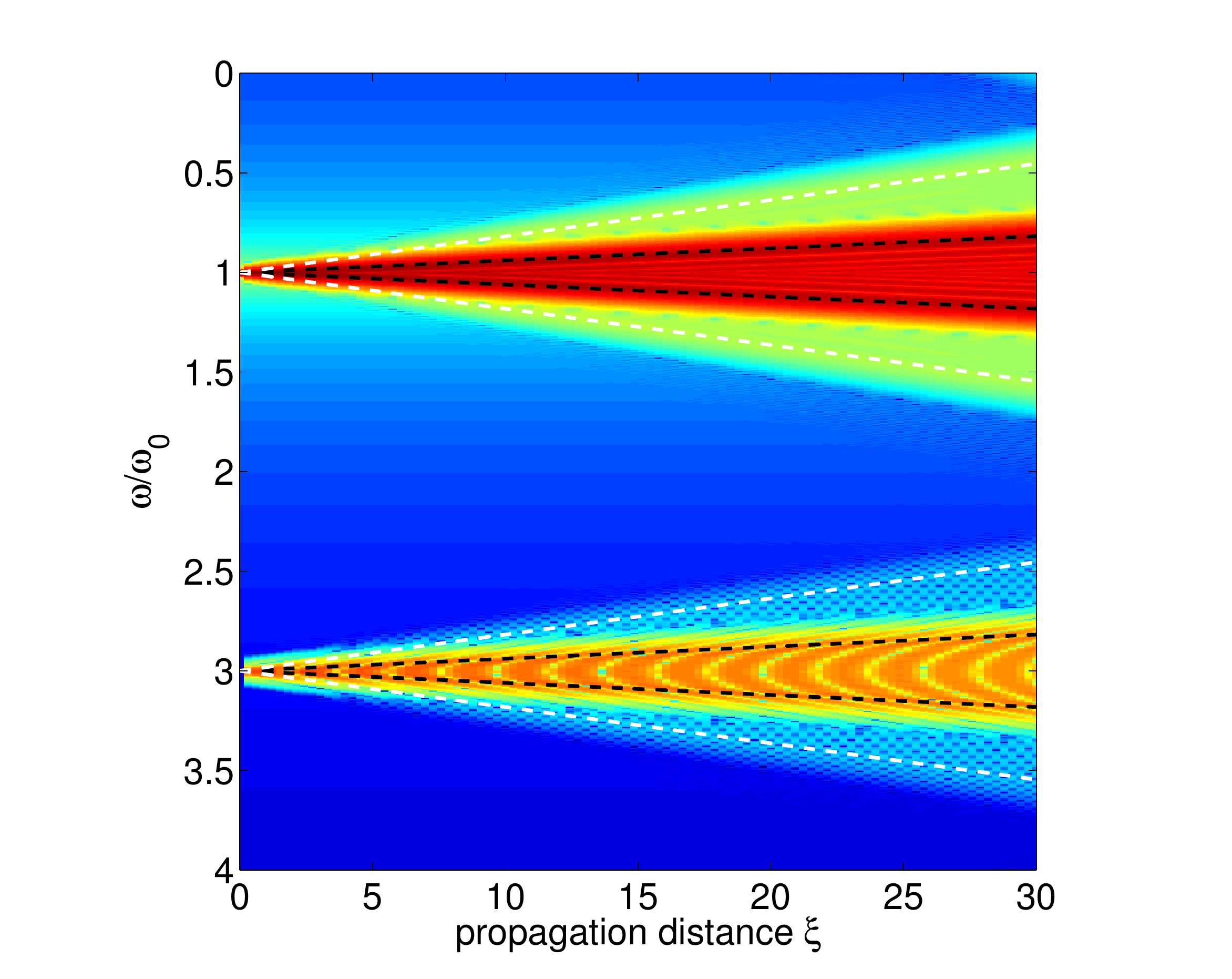}}
\caption{Contour plot of the spectral evolution of a Gaussian pulse $V=V_{0}\exp(-t^{2}/2)$. Parameters $\alpha$, $V_{0}$, $\kappa$ and $\mu$ are the same as in Fig. \ref{fig1}. Black dashed lines indicate the position of the conventional SPM lobes [$\Delta_{\pm}(\xi)=\pm(2/e)^{1/2}V_{0}^{2}\xi$], white dashed lines indicate the position of the new additional lobes [$\Delta'_{\pm}(\xi)\simeq3\Delta_{\pm}(\xi)$].} \label{fig3}
\end{figure}

In Fig. \ref{fig3}, a contour plot of the $\xi$-evolution of the spectrum is shown. The black dashed lines indicate the well known prediction for the old SPM lobes. For a Gaussian pulse $V=V_{0}e^{-t^{2}/2}$, this is given by $\Delta_{\pm}(\xi)=\pm(2/e)^{1/2}V_{0}^{2}\xi$, a formula that has been derived in several works, see for instance Ref. \cite{pinault}. For the new sidebands, since the presence of THG is an essential condition for their appearance, one expects that they accumulate approximately three times the nonlinear phase of the pump. This gives the prediction that the frequency detuning of the new SPM sidebands is approximately three times the detuning for the old sidebands, i.e. $\Delta'_{\pm}(\xi)\simeq\pm3\cdot(2/e)^{1/2}V_{0}^{2}\xi$. The white dashed line in Fig. \ref{fig3} shows the predicted position for the new sidebands. This is in excellent agreement with the spectral evolution determined by Eqs. (\ref{phinl1}). Note that both the fundamental frequency and the THG pulse undergo SPM and exhibit the formation of the additional sidebands.

The details of the spectra and the extent of the spectral broadening depend strictly on the particular pulse shaped used. We have found that for simple Gaussian pulses the visibility of the extra lobes and their amplitude is better than, for instance, hyperbolic secant and supergaussian pulses.

\begin{figure}[h!]
\centerline{\includegraphics[width=1\columnwidth]{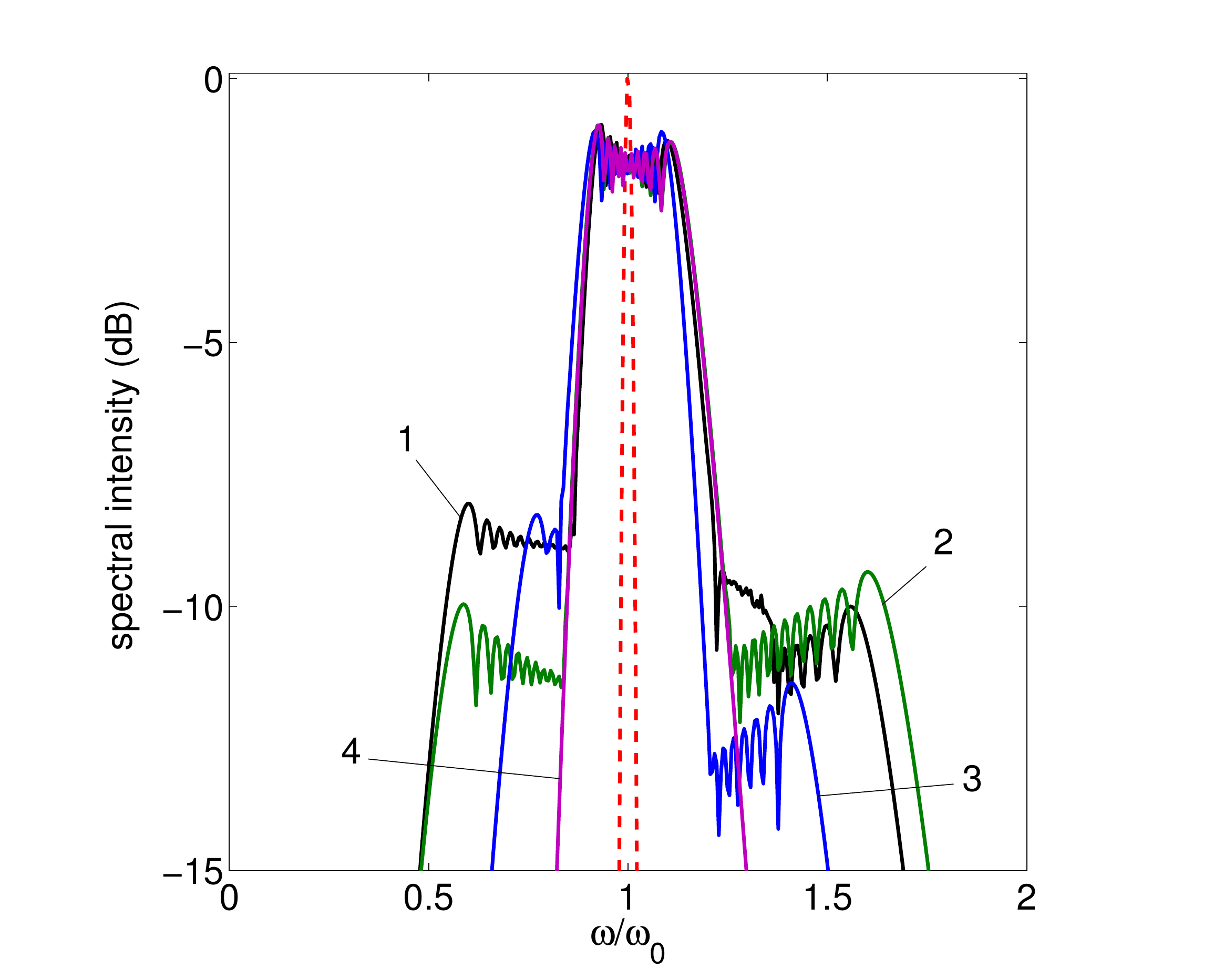}}
\caption{Simulation of the pulse propagation in diamond using the full UPPE. A comparison of the output spectra is shown when (1) all terms are included, (2) only THG is included, (3) the shock term is neglected and (4) all terms are switched off. One can see that the NFK greatly enhances the amplitude of the additional sidebands, while their position stays approximately the same. Physical parameters are $t_{0}=200$ fsec, $\lambda_{0}=1$ $\mu$m, $I_{0}=15$ TW/cm$^{2}$, $z=0.2$ mm. The red dashed line indicates the input pulse spectrum.} \label{fig4}
\end{figure}

In order to confirm our theoretical predictions based on Eq. (\ref{phinl1}), which assumes $V$ to be independent on $\xi$, we have performed numerical simulations based on the full UPPE. As a physically relevant and realistic example, we have used diamond, including its full dispersion and nonlinear polarization. In Fig. \ref{fig4} we show the spectral evolution of SPM for a fixed value of the propagation distance $z=0.2$ mm. The input pulse is a Gaussian with duration $t_{0}=200$ fsec and peak intensity $I_{0}=15$ TW/cm$^{2}$. The pump wavelength is $\lambda_{0}=1$ $\mu$m. A comparison of the output spectra is shown when (1) all terms are taken into account, (2) only THG is included, (3) we ignore the shock term and (4) all terms except the Kerr term are switched off. One can see that, as was predicted by Eq. (\ref{phinl1}), the NFK gives a massive contribution to the amplitude of the additional lobes. Moreover, one can observe that the shock term also contributes to the amplitude of the new radiation.


Looking at Fig. \ref{fig4} one can notice an asymmetry in the position and the amplitude of the additional lobes. This is due to dispersive effects (included in the UPPE) that were not taken into account in the calculation of the analytical nonlinear phase Eq. (\ref{phinl1}). The presence of the shock term also induces some asymmetry in the conventional SPM lobes, due to the characteristic blue-shift produced by the shock derivative \cite{shock}.



In conclusion, we have studied the formation of additional SPM lobes in the spectrum of an ultrashort and intense pulse induced by THG. We have found analytically the nonlinear phase in the approximation of constant pulse profile, which gives a qualitative understanding of the formation of the extra lobes. We have also solved numerically the UPPE in diamond to show that the extra lobes are formed in good agreement with the theory. We have elucidated the role of the different terms, including the shock operator and the negative frequency Kerr term, and found that they can dramatically enhance the amplitude of the lobes, when compared to the case when only the THG is included. This work will pave the way for the experimental demonstration of more nonlinear effects induced by negative frequency components of short pulses in nonlinear optics.

\end{document}